\documentclass[5p,fleqn]{elsarticle}
\usepackage[latin1]{inputenc}
\usepackage{graphicx}
\usepackage{hyperref}
\usepackage{breakurl}
\usepackage{upgreek}
\usepackage{lineno}

\journal{Solid-State Electronics}

\begin{document}

\title{Measurements of charge carrier mobilities and drift velocity saturation in bulk silicon of $<$111$>$ and $<$100$>$ crystal orientation at high electric fields}

\author[hh]{Julian Becker\corref{cor1}}
\ead{Julian.Becker@desy.de}
\author[hh]{Eckhart Fretwurst}
\author[hh]{Robert Klanner}
\cortext[cor1]{Corresponding author}
\address[hh]{Institute for Experimental Physics, University of Hamburg,\\ Luruper Chaussee 149, 22761 Hamburg}

\begin{abstract}
The mobility of electrons and holes in silicon depends on many parameters. Two of them are the electric field and the temperature. It has been observed previously that the mobility in the transition region between ohmic transport and saturation velocities is a function of the orientation of the crystal lattice.

This paper presents a new set of parameters for the mobility as function of temperature and electric field for $<$111$>$ and $<$100$>$ crystal orientation. These parameters are derived from time of flight measurements of drifting charge carriers in planar p$^+$nn$^+$ diodes in the temperature range between -30$^\circ$C and 50$^\circ$C and electric fields of 2$\times$10$^3$~V/cm to 2$\times$10$^4$~V/cm. 

\end{abstract}

\begin{keyword}
mobility \sep silicon \sep electric field \sep anisotropy
\end{keyword}

\maketitle

\setlength{\mathindent}{0mm}

\section{Introduction}
The transport properties in bulk silicon have been studied in great detail \cite{sverdlov}. 
Precise knowledge of the transport properties is important, as in applications with large drift distances, e.g. radiation sensors, even small uncertainties in the drift velocities can have noticeable effects.

In a radiation sensor ionizing radiation creates electron hole pairs in the silicon bulk material, which drift in the electric field created by an applied voltage and induce a signal in the readout electrodes\footnote{Unless a charge storing readout scheme is used (e.g. CCD- or DEPFET-sensors).}. Diffusion processes occur simultaneously causing a lateral spread of the charge carriers and a drift velocity distribution where some charge carriers are accelerated and some are decelerated due to the diffusion. The current induced by each charge carrier is proportional to its drift velocity, which, if diffusive processes are neglected, is given by the product of mobility and electric field. Thus, in order to optimize radiation sensors, precise knowledge of the mobility as function of the electric field is necessary.

Common radiation sensors are built in such a way that drift fields are in the order of 10$^4$~V/cm, making the charge collection times of these sensors sensitive to the charge carrier mobility in the transition regime between ohmic and saturation behavior. Especially for sensors damaged by irradiation the charge collection time is an important optimization parameters as charge carriers are trapped and thus degrade the obtained signal.

Parameterizations of charge carrier mobilities can be found in the literature, however the most recent measurements using the time of flight technique have been performed by Canali et al. in 1971 \cite{canali2}. There have been quite a few reviews of this data, the one by Jacoboni et al. \cite{jacoboni} is especially useful, as it provides a temperature depended parameterization of the drift velocity in $<$111$>$ direction.  There have been other reviews incorporating other measurements and proposing different parameterizations \cite{omar, selberherr, shukla, mohammad}, however either temperature dependence or anisotropy is neglected. For electric field values of 10$^4$~V/cm in $<$111$>$ orientated silicon at a temperature of 300~K the aforementioned parameterizations produce values scattered by approximately 10\% (best case), however the scattering increases significantly for different temperatures (57\% for holes at 250~K). None of the mentioned reviews provides a parameterization for the $<$100$>$ direction.

In this work measurements of pulse shapes of p$^+$nn$^+$ diodes in the temperature range between -30$^\circ$C and 50$^\circ$C with different applied bias voltages are compared to simulations in order to determine a set of parameters, describing the mobility as function of electric field and temperature for both $<$111$>$ and $<$100$>$ orientated silicon. 

\section{Time of flight method}
The time of flight method (sometimes also called transient current technique, TCT) is a well established technique to determine effective trapping times in irradiated silicon diodes \cite{kraner, kramberger}. It can also be used to investigate charge multiplication or effects of high charge carrier densities in segmented devices \cite{mikuz, becker}. Electron hole pairs are created inside a reverse biased diode either by ionizing radiation or laser light. These charge carriers then drift under the influence of the electric field and induce a current in the readout circuit. The current pulse contains information about the amount of drifting charge and their drift velocity. By using a laser source with appropriate wavelength the expected current pulses for x-rays, heavy ions or minimum ionizing particles can be simulated, allowing the optimization of application specific readout electronics. 

Planar p$^+$nn$^+$ silicon diodes with $<$111$>$ and $<$100$>$ crystal orientation were investigated using the time of flight method.

In order to measure the mobility of electrons and holes separately we create electron hole pairs by injecting laser light either from the p$^+$ or from the n$^+$ side. The penetration depth of the light is short compared to the thickness of the investigated samples, thus the other type of charge carriers is collected very quickly (within approximately 100~ps).

To evaluate whether effects of high charge carrier densities (also called plasma effects) were present the diameter of the laser spot on the sample was changed by focusing and defocussing. As no changes in the recorded current pulses were observed, high density effects can be considered negligible.

\section{Experimental setup}

\begin{figure}[tb]
  \includegraphics[width=0.5\textwidth]{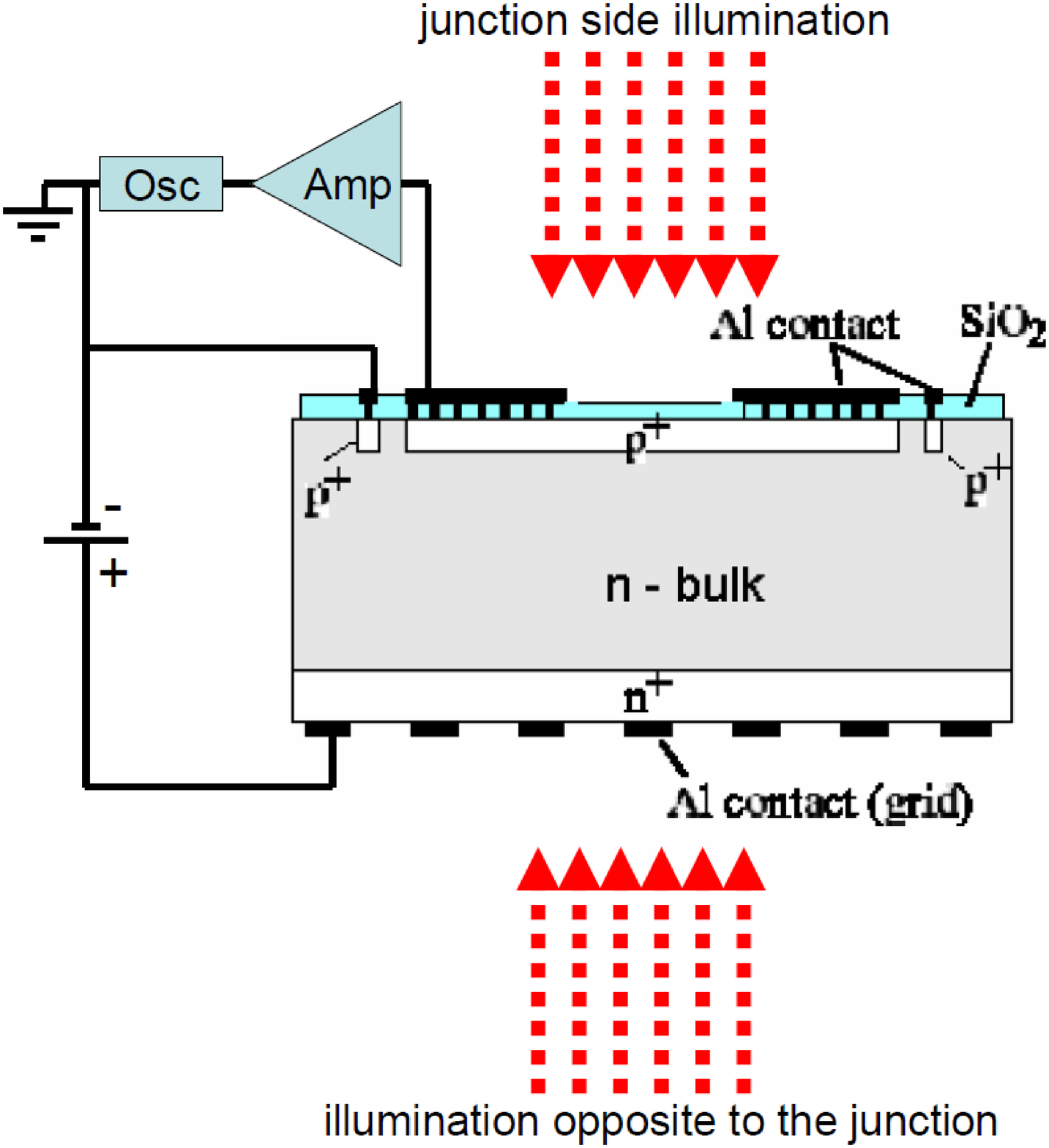}
  \centering
  \caption{Sketch of the experimental setup showing the polarity of the applied voltage. The laser light can be shined in either from the junction side or opposite to the junction (indicated by the arrows with broken lines). The diode has a thickness of 280~$\upmu$m or 272~$\upmu$m and is connected to the readout circuit.\label{diode}}
\end{figure}

Charge carriers were created with a laser of 660~nm wavelength and the time resolved current pulses of the investigated diode was read out by a Miteq AM-1309 wideband amplifier and a Tektronix DPO 7254 2.5~GHz oscilloscope. 

The sample was mounted on a substrate that allowed light injection from both sides while providing a stable ($\pm$ 0.1~K, rms) temperature in the range of 240~K to 340~K and applying the high voltage to the rear side of the diode. The guard ring was grounded.

\begin{figure}[tb]
  \includegraphics[width=0.5\textwidth]{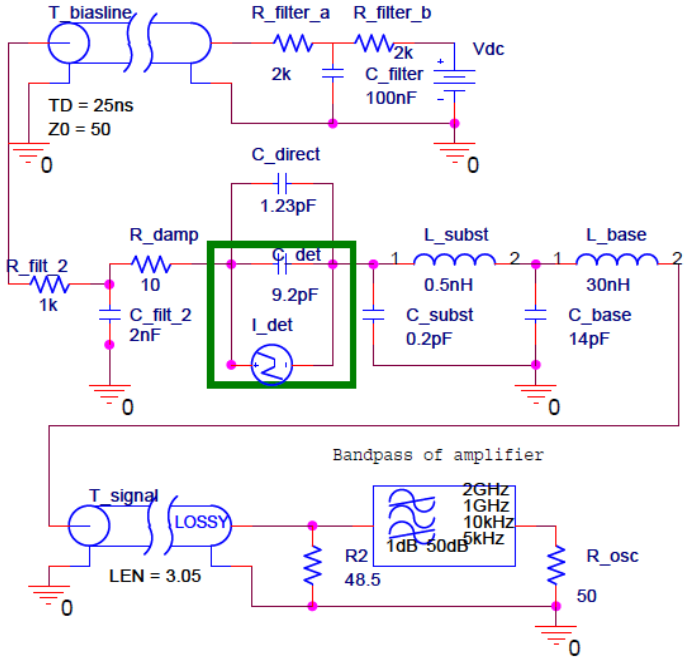}
  \centering
  \caption{Equivalent circuit for SPICE simulations. Dominant elements are the diode capacitance (only 9.2~pF shown) in parallel to the current source (marked by the rectangle), the system inductance (30~nH), the lossy transmission line (3.05~m RG-174 cable) and the band pass filtering properties of the amplifier.\label{circuit}}
\end{figure}

\subsection{Equivalent circuit for SPICE simulations}
\begin{figure}[tb]
  \includegraphics[width=0.5\textwidth]{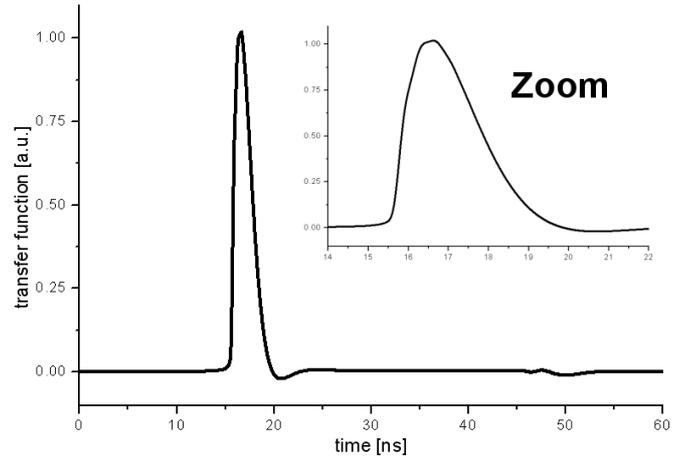}
  \centering
  \caption{Transfer function derived from the SPICE simulations for 9.2~pF capacitance. A similar transfer function was calculated for 9.4~pF capacitance, however the deviations are to small to be seen on this scale. The inset shows a zoom of the peak structure of the transfer function using the same units on the x- and y-axis.\label{transfer}}
\end{figure}

An equivalent circuit of the setup was used for SPICE simulations (see Figure \ref{circuit}). In the equivalent circuit detector capacitance (9.2~pF or 9.4~pF), signal cable length (3.05~m), bandwidth of the amplifier (1~GHz) and imperfections of the setup (inductances and capacitances) were taken into account. The transfer function of the whole setup has been checked and was found to introduce distortions, thus all simulations have been convoluted with the transfer function (shown in Figure \ref{transfer}).

\subsection{Laser properties}

The laser system was manufactured by Picoquant GmbH \cite{pq} and provided short and intense light pulses with a FWHM $<$~100~ps. The time resolved pulse structure was provided by the manufacturer. For this study laser light with a wavelength of 660~nm ($\pm$~2~nm) was used. The pulse energy of 140~pJ was attenuated with an optical attenuator, which has no effect on the time structure of the pulses. 660~nm light has an attenuation length of roughly 3~$\upmu$m in silicon at 20$^\circ$C.

\subsection{Investigated samples}

The investigated samples were planar p$^+$nn$^+$ diodes fabricated by CiS Forschungsinstitut f\"{u}r Mikrosensorik und Photovoltaik GmbH \cite{cis}. The silicon used is high resistivity n-type diffusion oxygenated float zone silicon with $<$100$>$ and $<$111$>$ orientation manufactured by Wacker.
The effective doping of the $<$100$>$ and $<$111$>$ samples are 8.2x10$^{11}$~cm$^{-3}$ and 1.1x10$^{12}$~cm$^{-3}$. The resulting depletion voltages are 49~V and 63~V with dark currents at the depletion voltage of 0.72~nA and 1.64~nA, respectively.
The p$^+$ implantation has a peak value of 3x10$^{19}$~cm$^{-3}$ and a junction depth of 0.5~$\upmu$m. The n$^+$ implantation has a peak value of 2x10$^{19}$~cm$^{-3}$ and depth of 1~$\upmu$m. The samples have very low concentrations of lattice defects, leading to trapping times in the order of milliseconds, which is many orders of magnitude larger than the longest pulse duration recorded in this work (approximately 35~ns). Thus any trapping effects have been considered negligible.

From the measured pad capacitance of 9.23~pF and the pad area of (4.95~mm)$^2$ we obtain a sample thickness of 280~$\upmu$m using the standard formula for a parallel plate capacitor without edge effects. The corresponding numbers for the $<$111$>$ diode are 9.42~pF, (4.93~mm)$^2$ and 272~$\upmu$m. For the measurement of the pad capacitance the capacitance of the guard ring to the backplane and the sensor edge is subtracted by the zero adjustment of the capacitance bridge. The remaining effect of the guard ring is estimated to be well below 1~\%. The estimated uncertainty of the diode thickness is $\pm$ 2~$\upmu$m.

Mechanical measurements of the thickness yield somewhat higher values but include 'dead' layers like passivation and aluminizations. The accuracy of the final capacitance has the biggest influence on the determination of the thickness, as a change of 0.1~pF changes the measured thickness by approximately 3~$\upmu$m.

The samples feature an opening on the top side and a grid like aluminum layer on the rear side to allow light injection. The gap between the metalization of diode guard ring is 20~$\upmu$m wide, the distance between the corresponding implantations is 10~$\upmu$m. A sketch of the sensor is shown in Figure \ref{diode}.

\section{Measurements}
Measurements have been performed for front side illumination (electron signal) and rear side illumination (hole signal). Bias voltages of 100~V, 200~V and 500~V were applied and these measurements were done in temperature steps of 10~K in the range between -30~$^\circ$C and 50~$^\circ$C (243.15~K and 323.15~K). A set of measurements and corresponding simulations are shown in Figure \ref{transients}.

To reduce the random noise component of the system averages 4096 pulses were taken. The remaining noise is no longer random and has an rms value of $<$~0.4~$\upmu$A.

\section{Simulations}

\begin{table*}[htb!]
	\centering
		\begin{tabular}{c|c|c|c|c|}
				& 	$\mu^0$ [cm$^2$/Vs]	&	$v^{sat}$ [cm/s]	& $\beta^{<100>}$ & $\beta^{<111>}$ \\
				\hline 
			electrons & 1440$\pm$15& (1.054$\pm$0.038)$\times$10$^7$ & 0.992$\pm$0.004& 1.075$\pm$0.004\\[0.1cm]
			holes 		&  474$\pm$10& (0.940$\pm$0.027)$\times$10$^7$ & 1.181$\pm$0.003& 0.924$\pm$0.002\\
		\end{tabular}
	\caption{Parameters for the field dependent mobility at 300~K.}
	\label{MobPars}
\end{table*}

\begin{table*}[htb!]
	\centering
		\begin{tabular}{c|c|c|c|c|}
				&	\multicolumn{4}{c}{Temperature parameter $\alpha_i$ for} \\
				& 	$\mu^0$ 	&	$v^{sat}$ 	& $\beta^{<100>}$ & $\beta^{<111>}$ \\
				\hline 
			electrons & -2.260$\pm$0.007& -0.602$\pm$0.003 & 0.572$\pm$0.003 & 0.220$\pm$0.001 \\[0.1cm]
			holes 		& -2.619$\pm$0.007& -0.226$\pm$0.002 & 0.644$\pm$0.003 & 0.550$\pm$0.002 \\
		\end{tabular}
	\caption{Temperature parameters $\alpha_i$ for the field dependent mobility reduction.}
	\label{TempPars}
\end{table*}

The simulations were done in two steps. At first the expected current pulse was calculated using a custom written transport code, then the resulting current pulse was convoluted with the time structure of the light pulse and with the transfer function of the setup.

The custom transport code is taking charge drift and diffusion into account. It is a 1D calculation on a grid with 100~nm cell size ($\Delta x$) with 5~ps time step ($\Delta t$).

Above full depletion the electric field inside the diode is linear with its highest value at the p$^+$n-junction ($x=0$) and described by the following equation:

\begin{equation}
	E(x)=\frac{1}{d}\left(U_{bias}+U_{dep}-\frac{2x}{d}U_{dep}\right)
\end{equation}

Where $x$ is the position inside the sensor (center of corresponding cell), $d$ the sensor thickness (280~$\upmu$m or 272~$\upmu$m), $U_{bias}$ the applied bias voltage and $U_{dep}$ the depletion voltage of the diode.

The normalized charge carrier distribution ($N_{e,h}(x,t=0)=\frac{\Delta x}{\lambda_{abs}}exp(-x^*/\lambda_{abs})$) generated by the laser light follows an exponential function with attenuation length $\lambda_{abs}$~= 3~$\upmu$m with $x^* = x$ for front side injection and $x^* = d-x$ for rear side injection.

The induced current is calculated according to the Ramo theorem \cite{ramo}. The weighting field of a planar diode is linear and the induced current (normalized to the elementary charge e$_0$) can be approximated as:

\begin{eqnarray}
	I(t+\Delta t) &=& \sum_x \left[I_{e}(x,t+\Delta t)-I_{h}(x,t+\Delta t)\right] \\
	I_{e}(x,t+\Delta t)&=& \frac{x e_0}{d\Delta t}\left(N_e(x,t+\Delta t)-N_e(x,t)\right) \\
	I_{h}(x,t+\Delta t)&=& \frac{x e_0}{d\Delta t}\left(N_h(x,t+\Delta t)-N_h(x,t)\right) 
\end{eqnarray}

\subsection{Drift and diffusion}

The charge drift and diffusion was simulated using the following basic equations:
\begin{eqnarray}
	\delta x(x) &=& \pm \overline{\mu_{e,h}}(x) \overline{E}(x) \Delta t \\
	\sigma^*_{e,h}(x) &=& \sqrt{2\overline{\mu_{e,h}}(x)\frac{kT}{e_0}\Delta t} \\
	N_{e,h}(x,t+\Delta t) &=& \sum_{x^*} N_{e,h}(x^*,t)C(x,x^*)\\
	C(x,x^*)&=& \frac{1}{2}\left[A(x,x^*)-B(x,x^*)\right] \\
	A(x,x^*)&=& erf\left(\frac{x^*+\delta x(x)-x+0.5*\Delta x}{\sqrt{2}\sigma^*_{e,h}(x)}\right)\\
	B(x,x^*)&=& erf\left(\frac{x^*+\delta x(x)-x-0.5*\Delta x}{\sqrt{2}\sigma^*_{e,h}(x)}\right)
\end{eqnarray}

Where $\delta x(x)$ is the distance traveled within one time step ($\Delta t$). It is negative for holes as they drift into the opposite direction of electrons. Average values (e.g. $\overline{\mu_{e,h}}(x)$) have been calculated by adding the values for the transversed cell and dividing by the number of cells traveled ($\pm 1/\delta x^* \sum_x^{x + \delta x^*}$, for $\delta x^*$ is $\delta x$ rounded to nearest cell center). $\delta x$ was determined by adding the drift times for each cell until the total drift time ($t_{drift}$) exceeded the time step when $n_{cell}$ cells have been transversed, then calculating $\delta x=\frac{n_{cell}*\Delta x}{t_{drift}}\Delta t$.
Diffusion is simulated using the Einstein relation ($D_{e,h}=\overline{\mu_{e,h}}(x)\frac{kT}{e_0}$). 

The number of charge carriers at each grid cell is calculated as the integral of a Gaussian function shifted by $\delta x$ with $\sigma = \sqrt{2D\Delta t}$ from $x-\frac{\Delta x}{2}$ to $x+\frac{\Delta x}{2}$ (beginning of cell to end of cell).

These shifted Gaussian functions are calculated for each grid cell and summed over all grid cells.

In order to avoid problems with charge carriers leaving the diode ($x<0$ or $x>d$) and diffusing back into the diode, a charge carrier which reaches the last grid point along the drift direction is removed from the simulation.

The numerical stability of the simulation has been checked by varying the grid size and time step by a factor of up to five. No effect on the pulse shape was observed. Charge conservation was checked and no charge loss was found.

\subsection{Mobility model}
The field dependence of the mobility was parameterized in the following way:

\begin{equation}
	\mu_{e,h}(x) = \frac{\mu_{e,h}^0} {\left(1+\left(\frac{\mu_{e,h}^0 E(x)} {v^{sat}_{e,h}}\right)^{\beta_{e,h}}\right)^{1/\beta_{e,h}}}
\end{equation}

Where $\mu_{e,h}(x)$ is the mobility of electrons or holes for a given grid cell $x$ with the electric field $E(x)$, $\mu_{e,h}^0$ the ohmic mobility (also called low field mobility), $v^{sat}_{e,h}$ the saturation velocity and $\beta_{e,h}$ a parameter describing the transition between ohmic behavior and saturation velocity.

The temperature dependence of these parameters was modeled using a simple exponential law:
\begin{equation}
	Parameter_i(T) = Parameter_i(300K) \left(\frac{T}{300K}\right)^{\alpha_i}
\end{equation}

\begin{figure*}[htb!]
  $\begin{array}{cc}
  \includegraphics[width=0.5\textwidth]{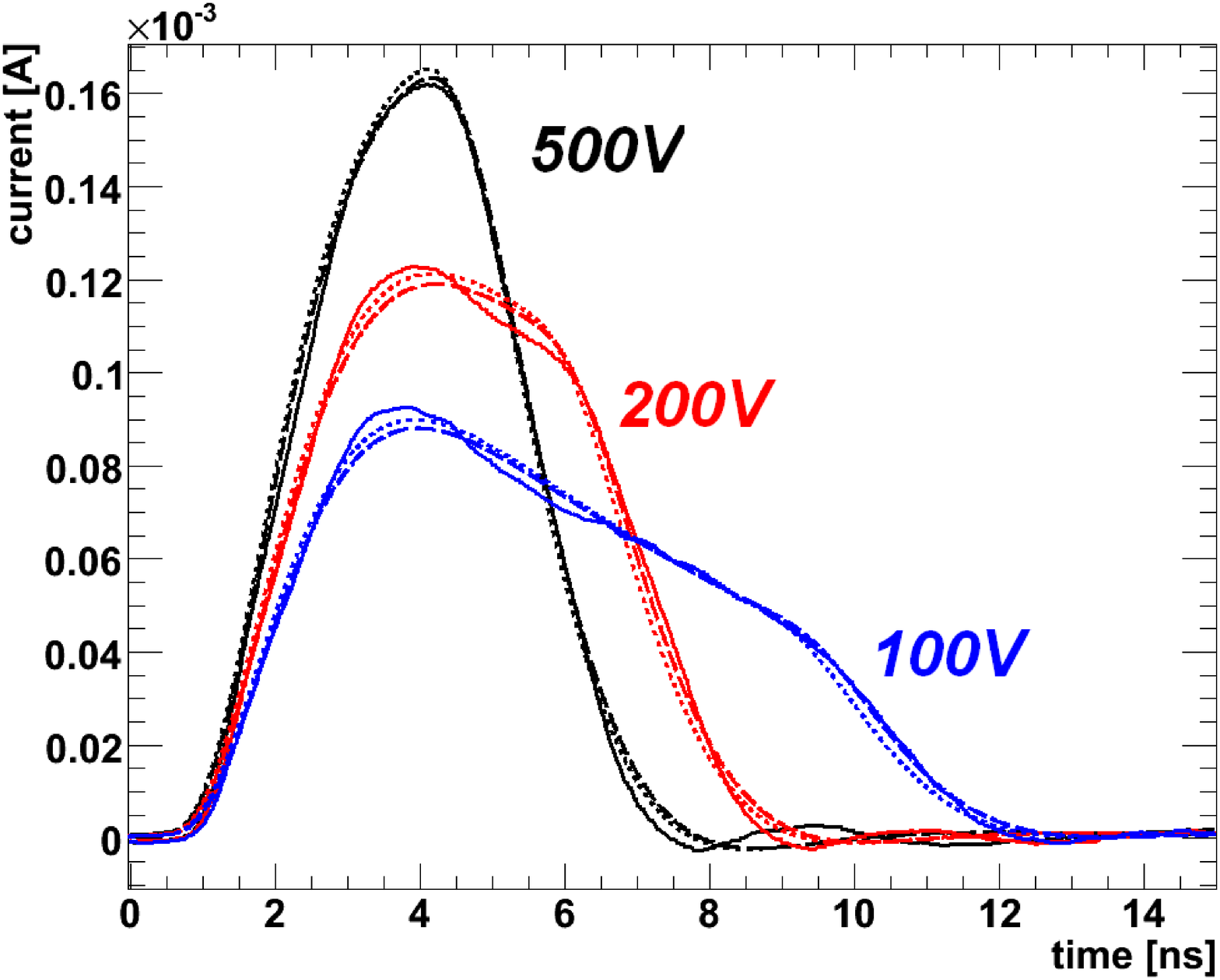} &
 	\includegraphics[width=0.5\textwidth]{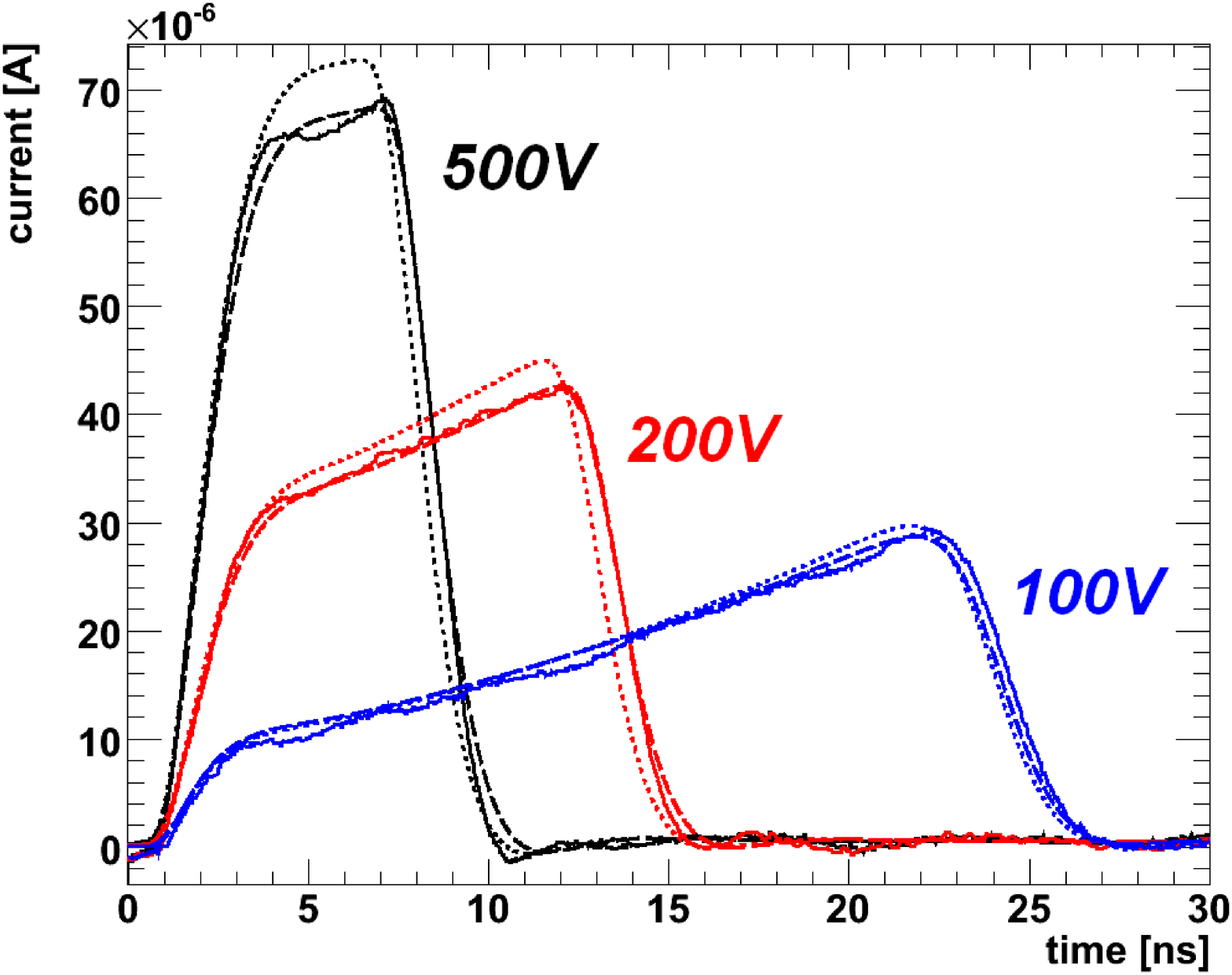} \\
  \includegraphics[width=0.5\textwidth]{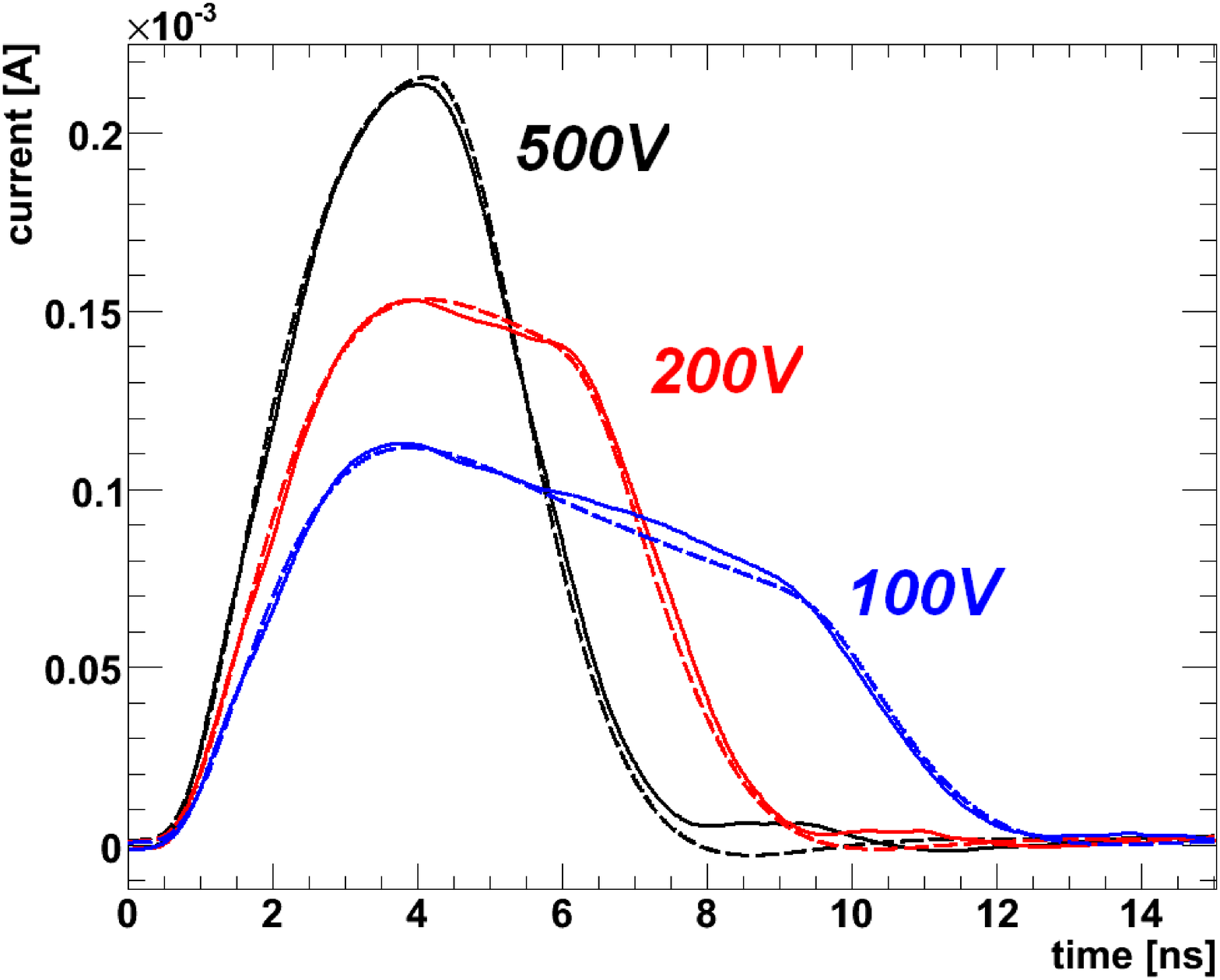} &
  \includegraphics[width=0.5\textwidth]{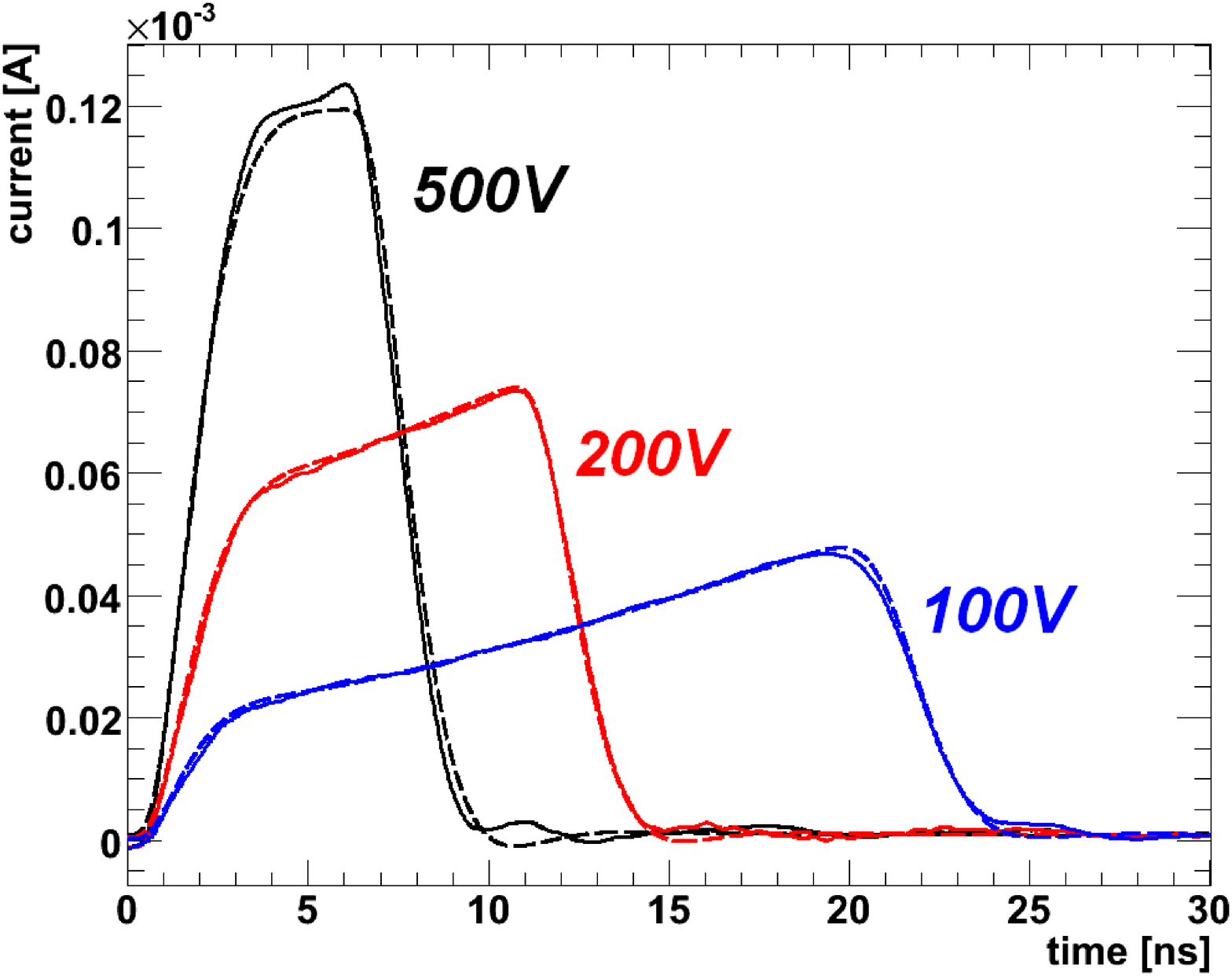} 
\end{array}$
  \centering
  \caption{Comparison of simulated pulses and measurements at T=30$^\circ$C. The upper row shows pulses for $<$111$>$ orientation, the lower for $<$100$>$ orientation; The left column electrons, the right column holes. The measured pulses (solid lines) are generally well reproduced by the simulations (dotted: using the mobility parameters of Jacoboni et al. \cite{jacoboni}, dashed: using the mobility parameters of this work), although some small ringing features are not correctly reproduced by the SPICE simulation. For electrons the simulated pulses of both models differ only slightly, while for holes a difference in pulse height and length is observed. \label{transients}}
\end{figure*}

\section{Results}

To determine the free parameters of the model a modified $\upchi^2$ procedure is used. At first current pulses for all measurements were simulated ($<$111$>$ and $<$100$>$ sample for 100~V, 200~V and 500~V at all temperature steps; 54 pulses in total) and the squares of the difference between measurement and simulation were summed up. This $\upchi^2$ sum was calculated for every set of parameters and the parameters with the smallest $\upchi^2$ are presented in Tables \ref{MobPars} and \ref{TempPars}. Errors have been calculated by using the Minuit fitting program \cite{minuit} of the ROOT package \cite{root}. Transients calculated with this set of parameters are shown in Figure \ref{transients}.

The errors given in Tables \ref{MobPars} and \ref{TempPars} have been determined by Minuit but should be seen as only approximate due to the high non-linearity of the system. 

To evaluate the influence of systematic uncertainties in the thickness and depletion voltage the same set of measurements was fitted, assuming $\pm$~1~\% difference in thickness or $\pm$~5~\% difference in depletion voltage. The difference in the resulting mobility parameters is within the range of the errors calculated by Minuit.

The penetration depth of 660~nm light changes by about 50~\% in the investigated temperature range. Comparison of simulations with the longest and shortest penetration depth and otherwise unchanged parameters showed no significant change in the overall pulse shape, thus the simulation neglected the temperature dependence of the penetration depth.

Optical photons of 660~nm wavelength (1.87~eV) produce so called hot charge carriers as the band gap of silicon is approximately 1.12~eV at room temperature. The simulation assumes charge carriers in thermal equilibrium with the crystal lattice. This is considered to be a good approximation as the thermalization of the hot carriers is usually very fast compared to the pulse duration.

\begin{figure}[tb]
  $\begin{array}{c}
  \includegraphics[width=0.5\textwidth]{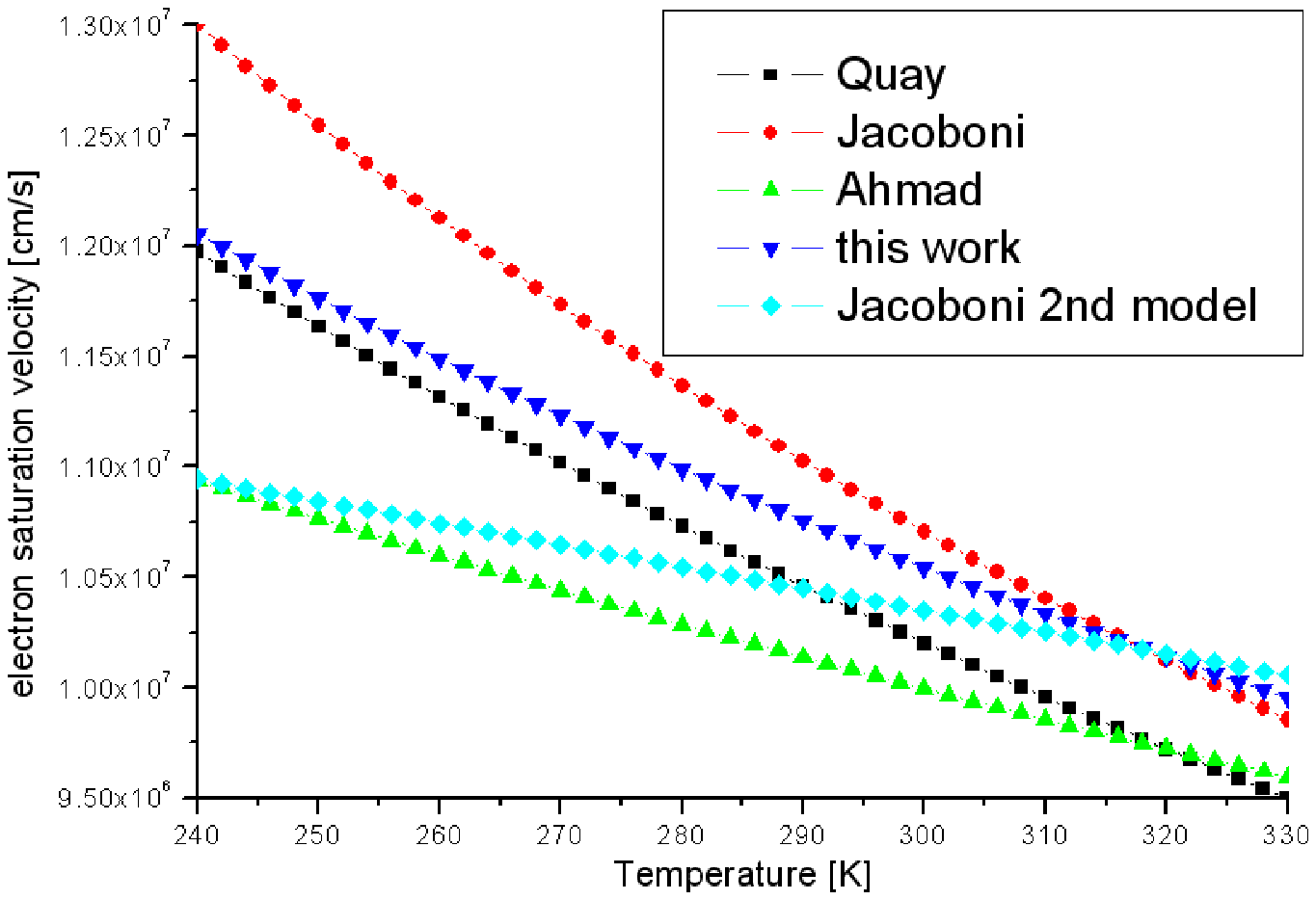} \\ 
  \includegraphics[width=0.5\textwidth]{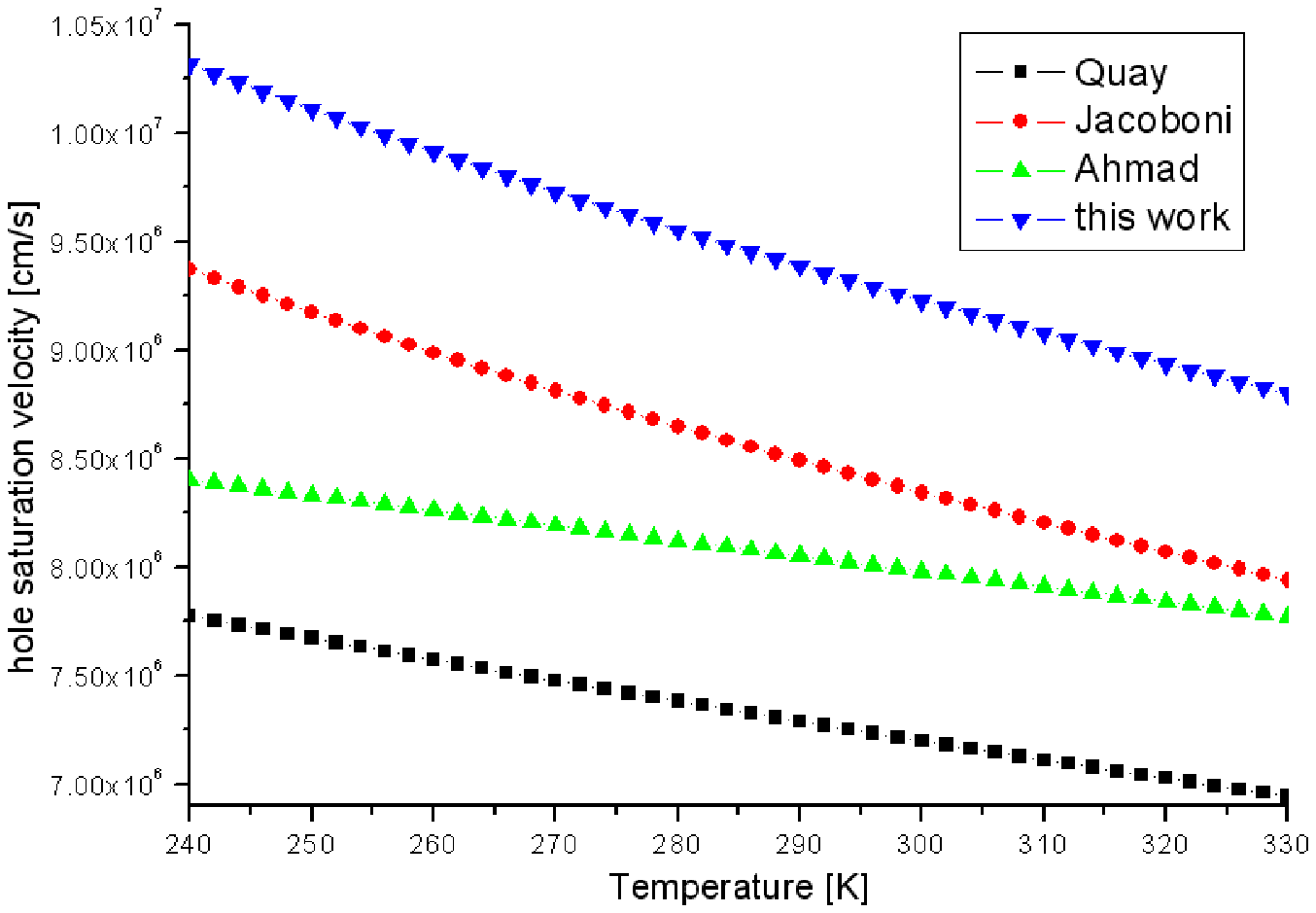}
\end{array}$
  \centering
  \caption{Comparison of different parameterizations of the saturation velocity proposed by other authors.  In the publication of Jacoboni et al. \cite{jacoboni} two parameterizations are given. The one parameterizing the electron saturation velocity directly has been labeled '2nd model'.\label{vsat}}
\end{figure}

\begin{figure}[tb]
  $\begin{array}{c}
  \includegraphics[width=0.5\textwidth]{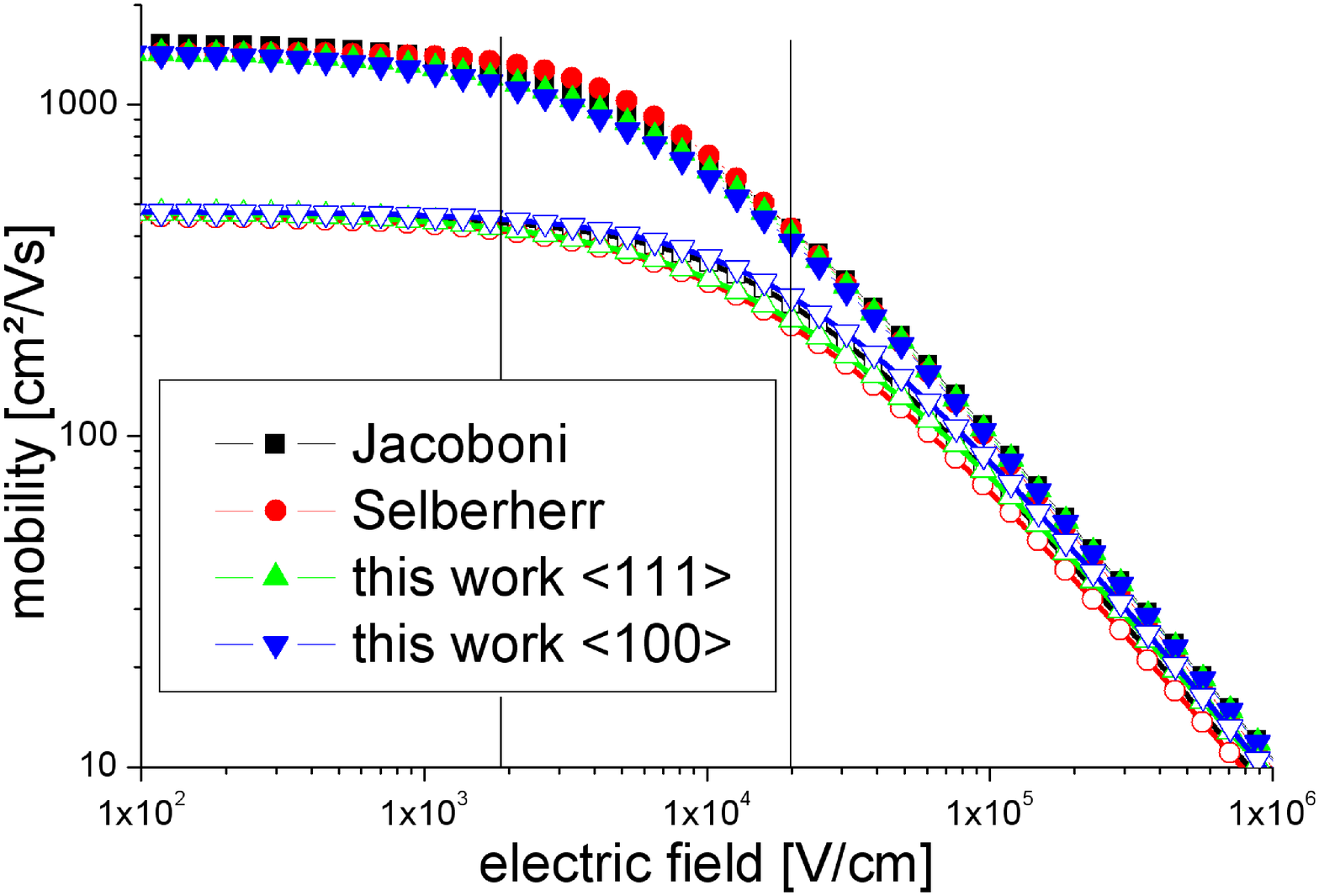} \\ 
  \includegraphics[width=0.5\textwidth]{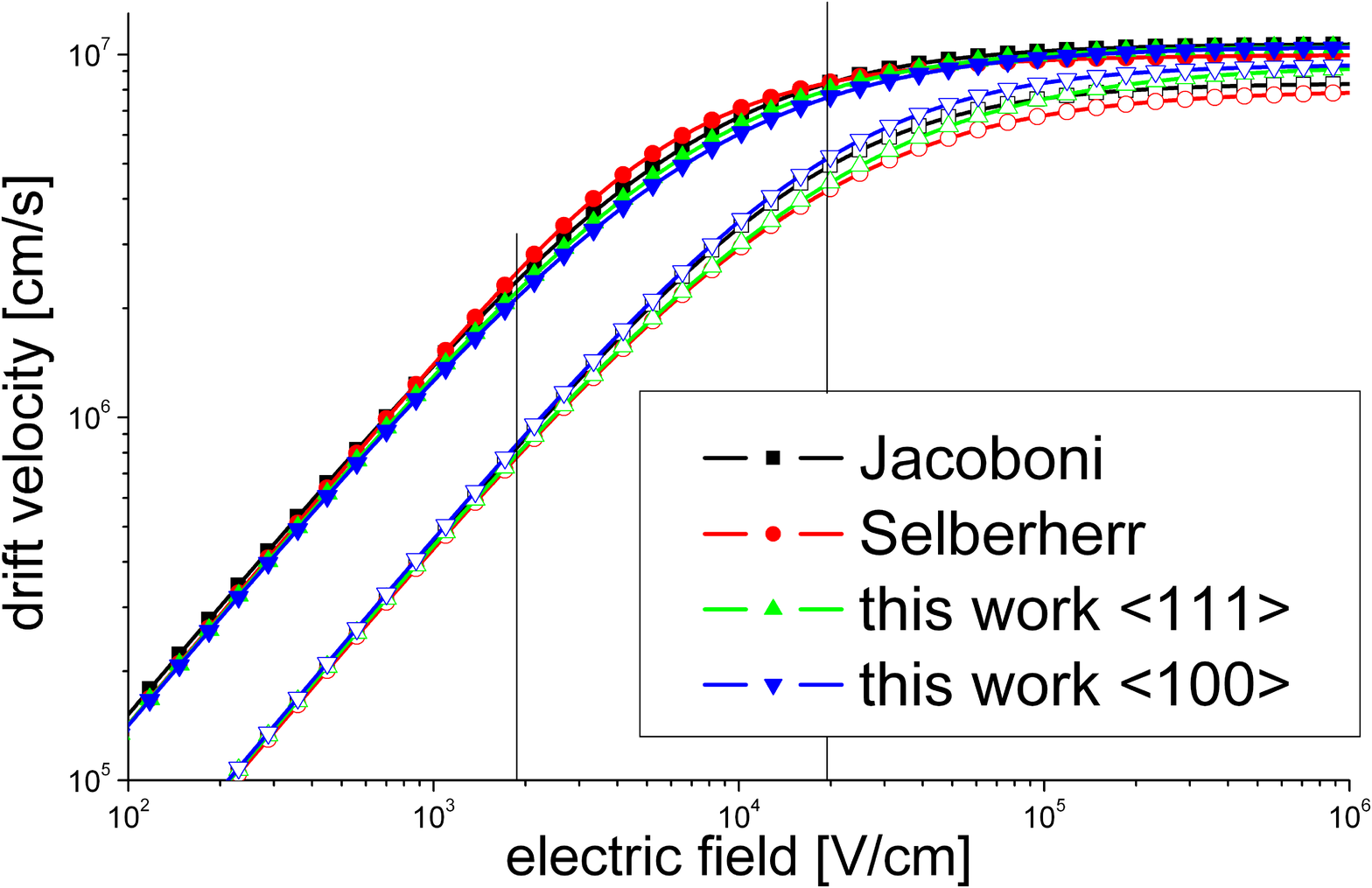}
\end{array}$
  \centering
  \caption{Comparison of different parameterizations of the mobility and drift velocities at T~=~300~K. Electrons are marked by the full symbols, holes by the open symbols. Vertical black lines indicate the approximate minimum and maximum field encountered in the measurements of this work. Both electrons and holes show an anisotropy. Comparing $<$111$>$ to $<$100$>$ orientation the electron drift velocity and mobility decreases, while the hole drift velocity and mobility increases.\label{mu}}
\end{figure}

\subsection{Fitted mobility parameters}
The fitted values of the ohmic mobilities at 300~K reproduce the accepted values \cite{green} within the quoted errors.
For the temperature coefficients reported values range from -2.3 to -2.6 for electrons and -2.2 to -2.8 for holes \cite{green}. The best fit value of holes lies within the quoted interval, while the electron value of this work lies just outside the quoted interval (rounded to the second digit it would be -2.3 and thus at the edge of the quoted interval), although it is lower than the value of -2.0 which has been reported by \cite{selberherr}.

Saturation velocity parameterizations as function of temperature have been provided by many authors \cite{jacoboni, ahmad, quay}, however the values differ by about 20~\%. The values for electrons of this work are in the range of values proposed by others, while the values for holes are higher than reported by \cite{jacoboni, ahmad, quay}, however values as high as 0.96$\times$10$^7$ cm/s at 300~K have been found by Smith and Frey \cite{smith}, which is even larger than the value determined in this work. A comparison is shown in Figure \ref{vsat}.

Except from the value of \cite{jacoboni} for $<$111$>$ orientation, no parameterizations of $\beta$ values as function of temperature and crystal orientation are available.

A comparison of mobilities and resulting drift velocities determined in this work with those proposed by other authors is shown in Figure \ref{mu}. A distinctive difference in the transition behavior between ohmic mobility and saturation velocity between the different parameterizations is observed. Especially for electrons a lower mobility than predicted in \cite{jacoboni, selberherr}is observed (for both crystal directions). In the transition region hole mobilities in the $<$111$>$ direction tend towards the values proposed by \cite{selberherr}, while the mobilities in the $<$100$>$ direction tend towards the values proposed by \cite{jacoboni}. In the saturation region the wide spread in the proposed velocities (shown in Figure \ref{vsat}) dominates, especially at lower temperatures (not shown).

\section{Discussion and Conclusions}

\begin{figure}[tb]
  \includegraphics[width=0.5\textwidth]{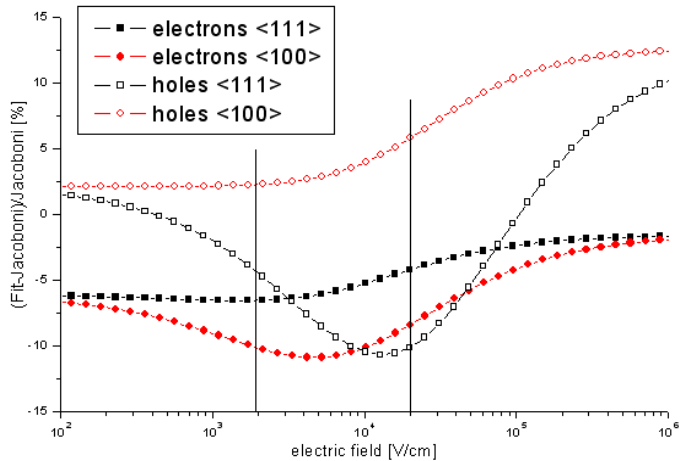} 
  \centering
  \caption{Difference between mobilities determined in this work and those of \cite{jacoboni} at T~=~300~K. Vertical black lines indicate the approximate minimum and maximum field encountered in the measurements of this work. Maximum deviations at this temperature are $\pm$~12\%.\label{mu_diff}}
\end{figure}

A new parameterization of the field and temperature dependence of electron and hole mobilities in bulk silicon was presented for $<$111$>$ and $<$100$>$ crystal orientation.

The differences between the values derived from the parameterization presented here and the parameterization of Jacoboni et al. \cite{jacoboni} are approximately $\pm$~15\% in the investigated temperature regime. The differences as function of electric field are shown in Figure \ref{mu_diff}. While the parameterizations for electrons in the $<$111$>$ show an almost constant difference of -~5\%, the other parameterizations show distinctive differences. The differences to the parameterizations proposed by others \cite{omar, selberherr, shukla, mohammad} are even larger and are not presented here.

No samples of $<$110$>$ orientation were available, thus no parameters for this direction could be determined. The electron parameters for this orientation provided by Castoldi and Rehak \cite{castoldi} ($\mu^0_e$~= 1394~cm$^2$/Vs, $v^{sat}_e$~= 1.85$\times$10$^6$~cm/s, $\beta^{<110>}_e$~=~2.51) differ greatly from the parameters presented here. However they have been obtained with very low fields (up to 700~V/cm), in which case the simple parameterization used in this work might need some corrections.

The presented results resolve the ambiguities in the temperature dependent mobility parameterizations for the $<$111$>$ orientation \cite{jacoboni, omar, selberherr, mohammad} and present a parameterization for the $<$100$>$ orientation. 
The results obtained enable more precise simulations of the charge collection in silicon sensors. The study is part of the investigation of the plasma effect, which occurs when the density of the electron hole pairs produced by the radiation exceeds the bulk doping density of the silicon. Its understanding is of relevance for the design of silicon pixel sensors for experiments at x-ray free-electron-lasers. It turned out, that the precision of the literature data on the transport parameters was insufficient to describe the reference data at low charge carrier densities. Therefore more precise data was required for a quantitative simulation of the plasma effect \cite{wias, thesis}.


\section{Acknowledgments}

This work was partly supported by the European XFEL, the Helmholtz Alliance ''Physics at the Terascale'' and the Federal Ministry of Education and Research (BMBF).

\end{document}